\documentstyle[sprocl,epsf]{article}

\input{psfig}

\def\Journal#1#2#3#4{{#1} {\bf #2}, #3 (#4)}

\def\PLB{Phys.~Lett.~B}
\def\PRL{Phys.~Rev.~Lett.}
\def\PRD{{Phys.~Rev.}~D}

\def\be{\begin{equation}}
\def\ee{\end{equation}}
\def\bea{\begin{eqnarray}}
\def\eea{\end{eqnarray}}

\begin{document}

\pagestyle{empty}

\begin{flushright}

FERMILAB-Conf-00/009-E\\
CDF/PUB/BOTTOM/CDFR/5204\\
\today
\end {flushright}

\vspace{0.5in}
\begin {center}
\begin{LARGE}
{\bf SPECTROSCOPY AND LIFETIME} \\
{\bf OF } \\
{\bf BOTTOM AND CHARM HADRONS } \\
\end{LARGE}

\begin {Large}
\vspace{10mm}
F{\normalsize UMIHIKO} U{\normalsize KEGAWA} \\
{\em Institute of Physics, University of Tsukuba \\
Tsukuba, Ibaraki 305-8571, Japan \\}
\vspace {5mm}
The CDF Collaboration

\vspace{5cm}
{\large
Invited talk at  the \\
              3rd International Conference on 
	$B$ Physics and $CP$ Violation \\
                       Taipei, Taiwan \\
                        December 3 -- 7, 1999 \\
}
\end {Large}
\end{center}

\newpage
\mbox { }
This is a blank page.
\newpage

\pagenumbering{arabic}          
\pagestyle{plain}

\title{
SPECTROSCOPY AND LIFETIMES 
OF BOTTOM 
AND CHARM HADRONS}

\author{Fumihiko Ukegawa}
\vspace*{3mm}
\author{(CDF Collaboration)}
\vspace*{3mm}
\address{
Institute of Physics, University of Tsukuba\\
Tsukuba, Ibaraki 305-8571 Japan\\E-mail: ukegawa@physics.px.tsukuba.ac.jp}


\maketitle\abstracts{ 
We review 
recent
experimental results 
on spectroscopy and lifetimes of 
bottom and charm hadrons.}
\section{Introduction}
There are several motivations for studying masses
and lifetimes of the hadrons
containing a heavy quark, either the bottom or the charm quark.
First, the mass and the lifetime are fundamental properties of
an elementary particle.
Second, the spectroscopy of hadrons
gives insights into the QCD potential between quarks.
In particular, a symmetry exists~\cite{HQS} 
for 
heavy  
hadrons 
when the heavy quark mass is taken to be infinite,
providing a powerful tool to predict and understand 
properties of those heavy hadrons.
Third, studies of the lifetimes of heavy hadrons 
probe their decay mechanisms. 
A measurement of the lifetime, 
or the total decay width,
is necessary when we extract 
magnitudes of elements of the Kobayashi-Maskawa matrix~\cite{CKM}.
Again, in the limit of an infinite heavy quark mass
things become simple and
the decay
of a heavy hadron should be the decay of the heavy quark $Q$.
This leads to a prediction that all hadrons containing the heavy quark $Q$
should have the same lifetime, that of the quark $Q$. 
This is far from reality in the case of charm hadrons, where
the $D^+$ meson lifetime is about 2.5 times longer than the $D^0$ 
meson lifetime. Perhaps the charm quark is not heavy enough.
The simple quark decay picture 
should be a better approximation 
for the bottom hadrons
because of the larger $b$ quark mass.

On the experimental side, the measurements and knowledge of the heavy
hadrons (in particular bottom hadrons)
have significantly 
improved over the last decade, thanks to
high statistics data accumulated by various experiments.
We shall review recent developments in these studies 
in the remainder of this manuscript.


\section{Charm Hadron Spectroscopy}\label{subsec:c_spec}

\begin{figure}[t]


\mbox{ \epsfxsize=0.50\textwidth 
	\epsffile{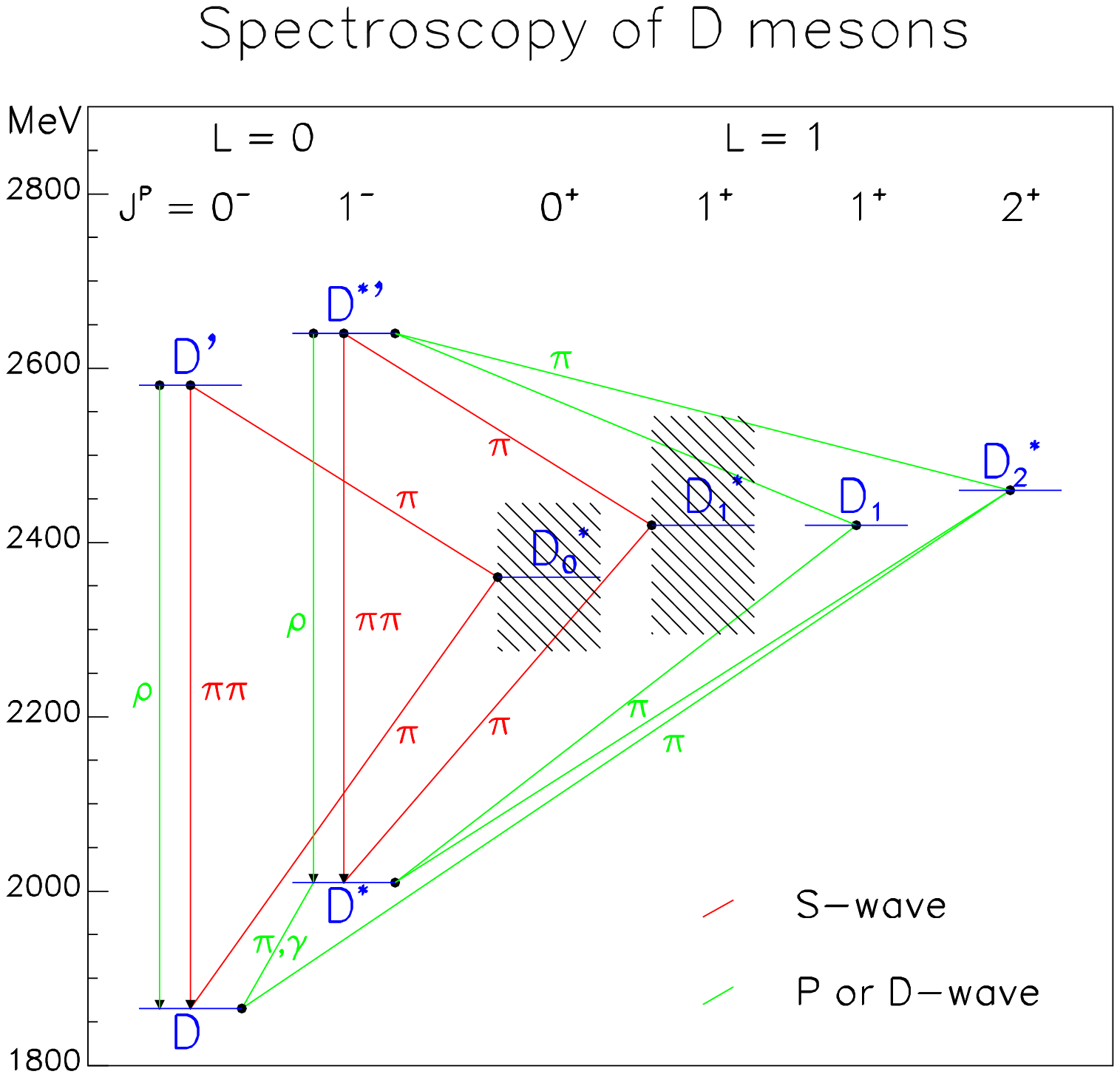}}
\mbox{ \epsfxsize=0.47\textwidth 
	\epsffile{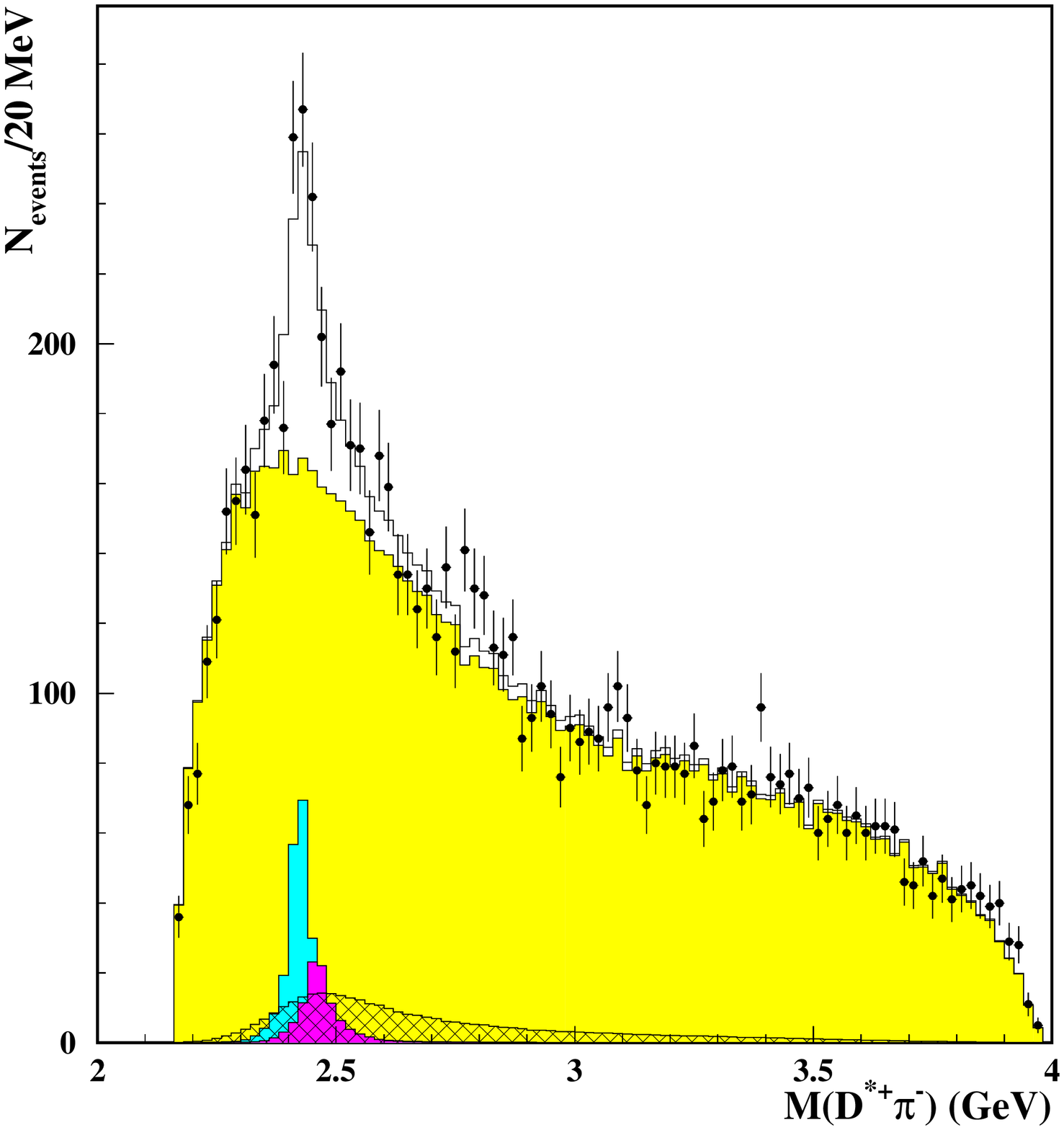}}
\vspace*{-5mm}
\caption{Left: Spectrum 
of charm mesons.
Right: $D^{*+} \pi^-$ mass spectrum by CLEO.
}
\label{fig:charm_spec}
\end{figure}

Let us consider 
mesons 
consisting of a charm quark $c$ and a light antiquark $\bar q$
(Figure~\ref{fig:charm_spec}(left)).
The ground states 
are the pseudoscalar ($D$) and
the vector ($D^*$) mesons, 
which  have long been established.
They have zero orbital angular momentum ($L=0$) between the quarks,
and the total angular momentum of the mesons are given by addition
of two spin 1/2 particles, namely $J=S=0$ or 1. 

If we allow for one unit of orbital angular momentum ($L=1$)
and combine it  with the total spin 
$S$ of the $c \bar q$ system,
there will be 
four states, $J^P = 1^+$, $0^+$, $1^+$ and $2^+$.
These states are sometimes called the $D^{**}$ mesons.
They are expected to decay predominantly to $D^{(*)} \pi$ 
pairs via the strong interaction.
Angular momentum and parity conservation 
restrict 
the decay of the $0^+$ meson to the $D \pi$ final state.
Similarly, 
the $1^+$ meson decays only to $D^* \pi$ pairs.
The $2^+$ meson can decay either to $D^* \pi$ or $D \pi$.

In the limit of an infinite heavy quark mass,
$m_Q = \infty$, 
the light degree of freedom is decoupled from the heavy quark, 
and 
the total angular momentum
of the light antiquark $j = L + s_{\bar q}$ 
is a good quantum number. 
In this case 
the four states above form
two doublets, each having $j=1/2$ ($J=0,1$)
or $j=3/2$ ($J=1,2$), 
where two members of each doublet 
are degenerate in mass, width and other quantum numbers.
The decay of the $j=3/2$ doublet proceeds through a $D$-wave,
while
the decay of the $j=1/2$ doublet proceeds through an $S$-wave.
Therefore, the $j=3/2$ doublet is expected to be narrow
and the $j=1/2$ doublet to be wide.

The two narrow states, 
$D_1$ and $D_2^*$ mesons in the PDG notations, 
have been established~\cite{PDG}.
Here we describe the first observation of a wide state by 
the CLEO experiment~\cite{d2st_wide}.
CLEO uses 3.1 
fb$^{-1}$ of $e^+ e^-$ annihilations
at the $\Upsilon(4S)$ resonance. 
The $B^- \rightarrow D^{*+} \pi^- \pi^-$ decay
is used, and the $D^{*+} \pi^-$ pairs are studied.
The $D^{*+}$ meson is reconstructed in the $D^0 \pi^+$ decay mode.
The usage of this decay chain gives enough constraints 
that allow the study of the $D^{*+} \pi^-$ pairs
without explicitly reconstructing the $D^0$ meson,
resulting in large 
increase in statistics.
In particular, 
the angular correlations that
arise from the fact that the $D^{**}$ meson
is fully polarized in the decay
are exploited intensively.
Figure~\ref{fig:charm_spec}(right) shows the 
$D^{*+} \pi^-$ mass spectrum. 
The background subtracted distributions are also shown, 
as are the two narrow states, 
as well as 
a wide component (hatched histogram).
The mass and width of the wide component
are measured to be
$m = 2461 \, ^{+\, 41}_{-\, 34} \pm 10 \pm 32$ MeV/$c^2$ 
and
$\Gamma \,  = 290 \, ^{+\, 101}_{- \,\,79} \pm 26 \pm 36$ MeV/$c^2$.
Also the branching fraction is measured to be
${\cal B} ( B^- \rightarrow D^{*0}_1 \pi^-)  \cdot
 {\cal B} ( D^{*0}_1 \rightarrow D^{*+} \pi^- ) 
= 
 ( 10.6 \pm 1.9 \pm 1.7 \pm 2.3) \times 10^{-4}$.

The DELPHI experiment reported~\cite{delphi_radial} 
evidence for a radial excitation 
state $D^*$$'^+$ using the $D^{*+} \pi^+ \pi^-$ final state.
However, neither OPAL nor CLEO 
has been able to confirm it~\cite{radial_no}.

CLEO has found~\cite{cleo_charm_baryon} 
a new charmed baryon state that decays
through $\Xi_c^* \pi$ to  $\Xi_c \pi^+ \pi^-$.
Both charged and neutral states are observed
as a peak in the mass difference 
$\Delta m \equiv m(\Xi_c \pi^+ \pi^-) - m(\Xi_c)$,
at 
$348.6 \pm 0.6$ (charged) and $347.2 \pm 0.7$ MeV/$c^2$ 
(neutral).
An upper limit on the width is placed to be
$\Gamma < 3.5$ (charged) and $< 6.5$ (neutral) MeV/$c^2$.
They are interpreted as the $J^P = \frac{3}{2}^-$ $\Xi_{c1}$
isospin doublet, namely the $L=1$ orbital excitation of the $\Xi_c$ baryon.

\section{Bottom Hadron Spectroscopy}\label{subsec:b_spec}
A lot of progress has been made  in the spectroscopy of
bottom hadrons in the past few years as well.
All ground states have been established, and even the
$B_c^-$ meson, a bound state of the two different kinds of
heavy quarks, has been 
observed~\cite{b_c}.
The $B^*$ mesons have also been established, at a mass about 50 MeV/$c^2$
above $B$. 
The mass splitting is smaller than in the charm mesons 
($\sim 150$ MeV/$c^2$), 
certainly consistent with expectations
from the heavy quark symmetry.

\begin{figure}[t]


\mbox{ \epsfxsize=0.47\textwidth 
	\epsffile{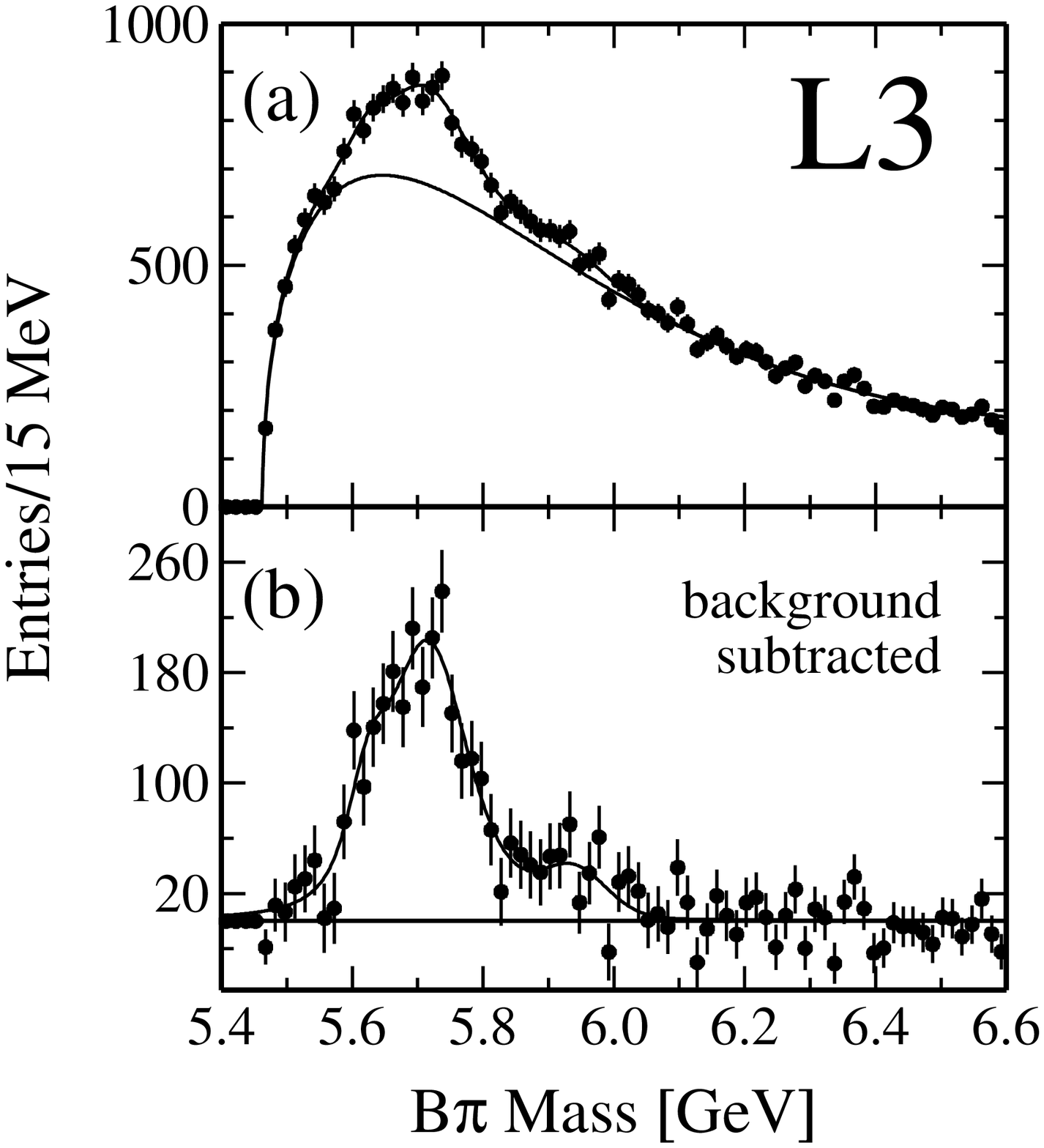}}
\vspace{-3mm}
\mbox{ \epsfxsize=0.5\textwidth 
	\epsffile{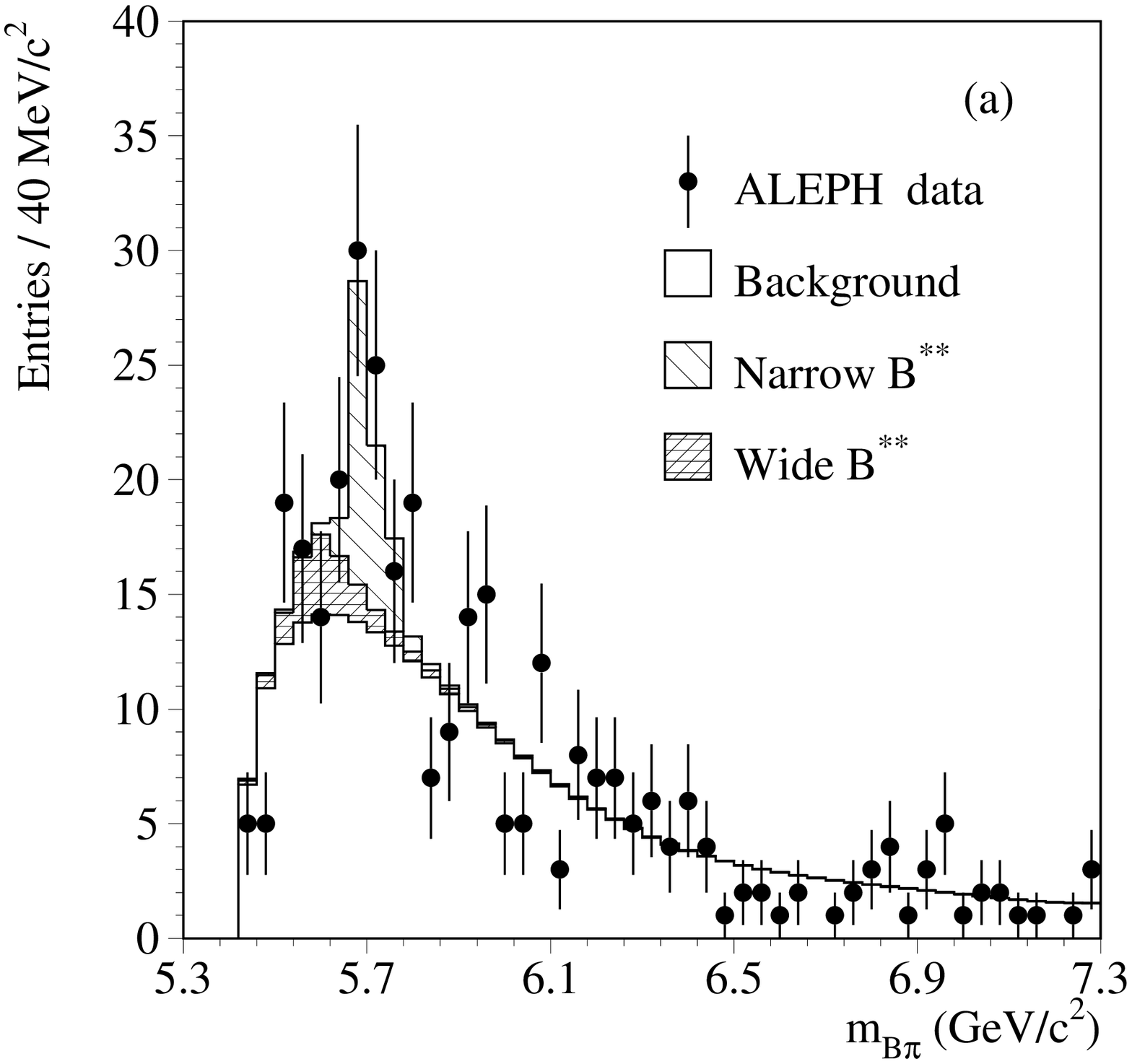}}
\caption{
$B^{**}$ analyses by L3 (Left) and ALEPH (Right).
$B \pi$ mass spectra are shown. 
}
\label{fig:L3_b2st}
\end{figure}

The $B^{**}$ states have also been observed.
Here we describe an L3 analysis~\cite{L3_b2st}
as a recent example.
The $B$ mesons are reconstructed inclusively using secondary vertices
and are 
combined with pion candidates coming from the primary vertex.
Figure~\ref{fig:L3_b2st}(left) shows the $B \pi$ mass distribution,
where an enhancement is observed around 5.7 GeV/$c^2$.
A fit for the masses and widths of the narrow and the wide
states is performed, assuming, 
within each doublet,
the mass splitting of 12 MeV/$c^2$
and  equal widths,
and relative production rates
according to $2J+1$ spin counting.
This yields
$m = 5768 \pm 5 \pm 6$ MeV/$c^2$ and $\Gamma = 24 \pm 19 \pm 24$ MeV/$c^2$
for the $B^*_2$ state, 
and
$m = 5670 \pm 10 \pm 13$ MeV/$c^2$and $\Gamma = 70 \pm 21 \pm 25$ MeV/$c^2$
for the $B^*_1$ state.
The fraction of the $B^{**}$ production is extracted to be
$f( b \rightarrow B^{**} \rightarrow B^{(*)} \pi) = 0.32 \pm 0.03 \pm 0.06$.
The fit actually includes a fifth component, 
radial excitation states $B'$, 
to account for an enhancement near 5.9 GeV, 
giving 
$m = 5937 \pm 21 \pm 4$ MeV/$c^2$,
$\sigma = 50 \pm 22 \pm 5$ MeV/$c^2$, and
$f( b \rightarrow B' \rightarrow B^{(*)} \pi) = 0.034 \pm 0.011 \pm 0.008$.

Although the $B^{**}$ mesons are collectively observed,
their individual states have not been separated. 
This is because (a) typical $B$ reconstruction 
relies on inclusive techniques and thus suffers from
a relatively poor mass resolution,
(b) the photon in the $B^* \rightarrow B \gamma$ decay
is not usually included and thus $B^* \pi$ and $B \pi$ states
are not separated, and (c) all $B$ hadron species are mixed
together.

OPAL~\cite{opal_b2st} attempted to remedy the point (b) above, 
with a statistical separation of $B^* \pi$ and $B\pi$ states
by looking for the photon from the $B^*$ decay. 
Inclusively reconstructed $B$ decay events are separated into
$B^*$-enriched and $B^*$-depleted samples, and the $B^{**}$ branching fraction
of ${\cal B} ( B^{**} \rightarrow B^* \pi) = 
0.85 \, ^{+\, 0.26} _{-\, 0.27} \pm 0.12$
is extracted.

ALEPH~\cite{aleph_b2st} uses fully reconstructed decays
$B \rightarrow {\overline D}^{(*)} n\pi$
and $J/\psi K^{(*)}$, 
solving difficulties (a) and (c). 
Signals of about 200 $B^+$ and 140 $B^0$ are observed.
The $B\pi$ mass spectrum is shown in Figure~\ref{fig:L3_b2st}(right).
The limited statistics, however, prevent the determination of the masses and
widths of all four states separately. A fit similar to that in
the L3 analysis is performed, assuming 
a mass splitting within a doublet of 12 MeV/$c^2$, 
a mass splitting between the narrow and wide doublets  of 100 MeV/$c^2$,
and 
widths
$\Gamma(B_2^*) = 25$ MeV/$c^2$, 
$\Gamma(B_1) = 21$ MeV/$c^2$ and $\Gamma = 150$ MeV/$c^2$ for wide states,
as well as relative production rates according to $2J+1$ spin counting.
The fit yields
$m(B_2^*) = 5739 \, ^{+\, 8} _{-\, 11} \, ^{+\, 6} _{-\, 4}$ MeV/$c^2$
and 
$f( b \rightarrow B^{**} \rightarrow B^{(*)} \pi) = 0.31 \pm 0.09 
\, ^{+\, 0.06} _{-\, 0.05}$.
CDF~\cite{cdf_b2st} performs a similar analysis using semileptonic
decays ${\overline B} \rightarrow \ell^- \bar \nu D^{(*)} X$.

There is one important application of 
$B^{**}$ mesons for other $B$ physics studies, namely 
flavor tagging.
It is an essential ingredient 
for measurements of $B^0 {\overline B}$$^0$ oscillations
and some $CP$ violation phenomena. 
It has been customary to infer the flavor of the $B$ hadron
of interest 
by identifying the flavor of the other
$B$ hadron in the event, using the fact that $b$ and $\bar b$ quarks
are produced in pairs.
It is also possible to exploit the charge-flavor correlation
between the $B$ hadron and the nearby charged pion~\cite{SST_th}.
A well known example would be the decay $D^{*+} \rightarrow D^0 \pi^+$
when you look for 
$D^0 {\overline D}$$^0$ mixing.
If it is accompanied by 
$\pi^+$, 
it must have been produced as $D^0$, 
not ${\overline D}$$^0$. 
The $B^*$ mesons cannot produce a pion kinematically. 
Therefore, $B^{**}$ mesons are the main resonant state
that can produce correlated pions. 
They
can also be produced in the fragmentation (non-resonant) 
processes. The charge correlations are the same 
whether it is resonant or non-resonant.
The method has been applied successfully to
$B^0 {\overline B}$$^0$ oscillation measurements~\cite{SST}.
The ALEPH analysis above also performs tagging studies
and finds similar tagging effectiveness.
Tag purity could be improved further 
if the $B\pi$ mass resonant regions are selected.


\section{Charm Hadron Lifetimes}\label{subsec:c_life}

Recent developments in charm lifetime measurements 
include (a) a new precision in the $D^+_s$ lifetime (E791, CLEO)
and (b) startup of new experiments (FOCUS, SELEX).
Here we describe the $D_s^+$ lifetime measurements.

\begin{figure}[t]
\mbox{ \epsfxsize=0.40\textwidth 
	\epsffile{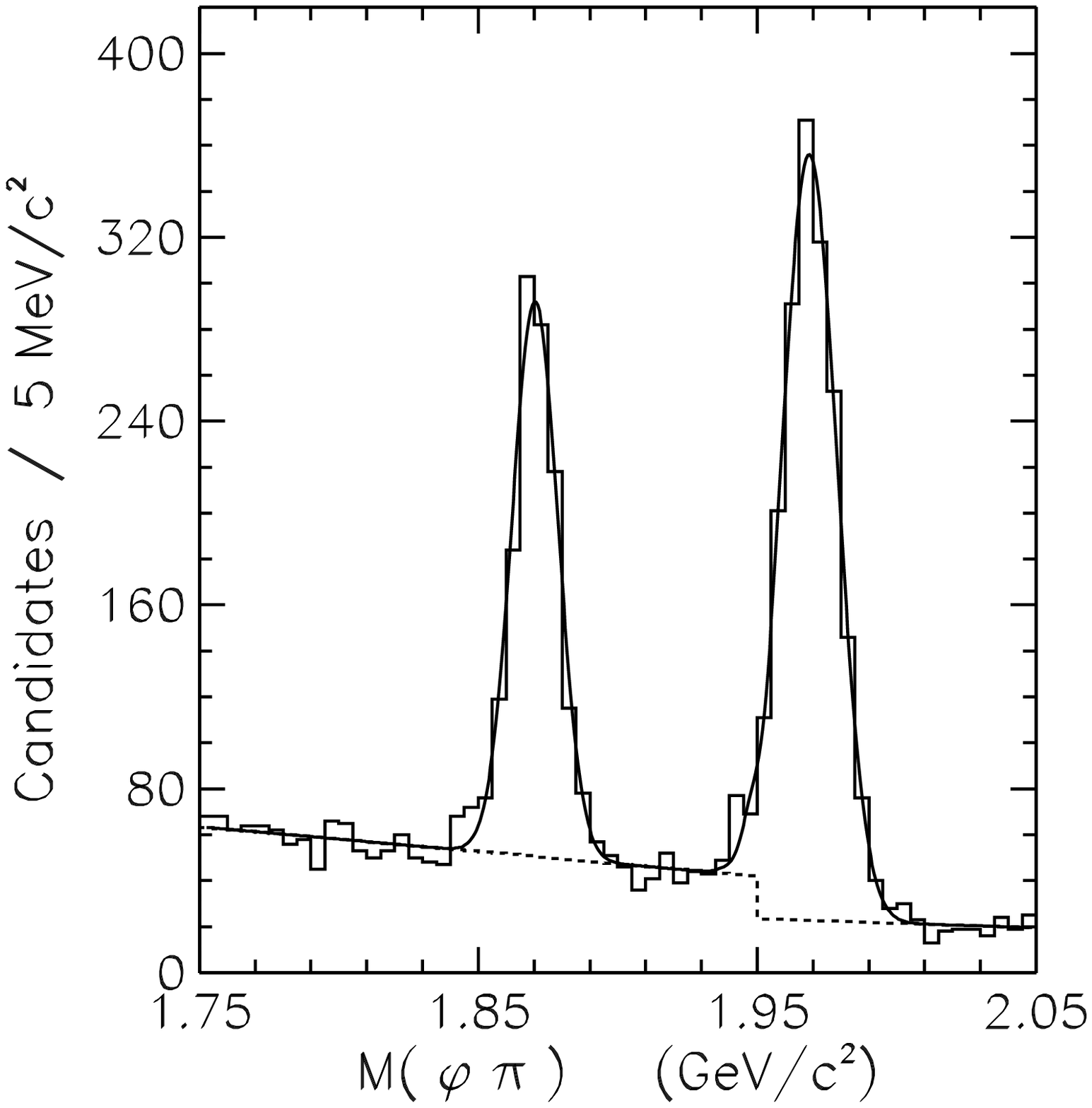}}
\vspace{-4mm}
\mbox{ \epsfxsize=0.55\textwidth 
	\epsffile{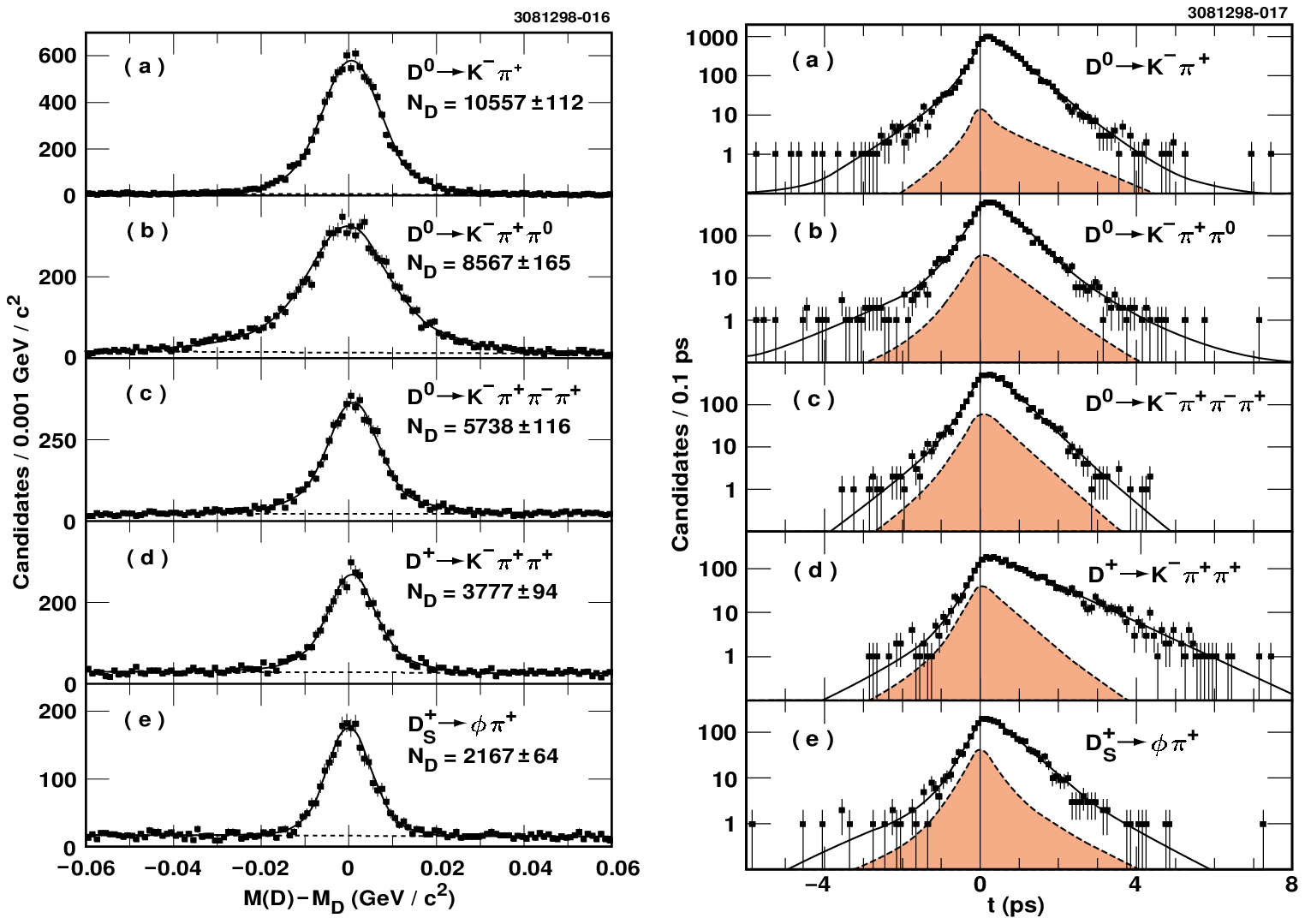}}

\caption{
Left: $D^+_s$ lifetime measurements by E791.
Right: Charm lifetime measurements by CLEO.
}
\label{fig:ds_791}
\end{figure}

E791~\cite{e791} uses data from 1991 - 92 
with 500 GeV/$c$ $\pi^-$ beam on foil targets.
The $D^+_s$ meson is reconstructed with the $\phi \pi^+$, 
$\phi \rightarrow K^+ K^-$ mode.  
Figure~\ref{fig:ds_791}(left)
shows the $\phi \pi^+$ mass spectrum. A signal of $1662 \pm 56$
is estimated. 
The peak around 1860 MeV/$c^2$ is from the Cabibbo-suppressed $D^+$ decay. 
The dashed line represents combinatorial background, 
where a discontinuity arises because 
the $D_s^+ \rightarrow \phi \pi^+$ candidates were
required to be inconsistent with
the $D^+ \rightarrow K^+ \pi^- \pi^+$ 
hypothesis.
Mass and decay time distributions are fit simultaneously 
and  the $D_s^+$ lifetime is extracted to be 
$\tau (D_s^+) = 518 \pm 14 \pm 7$~fs. Using the $D^0$ lifetime
of $415 \pm 4$~fs (PDG 1998), it is found that 
$\tau (D_s^+) / \tau(D^0) = 1.25 \pm 0.04$.

CLEO~\cite{ds_cleo} uses 3.7~fb$^{-1}$ of $e^+ e^-$ annihilation
near the $\Upsilon(4S)$. The $D^+_s$, as well as $D^0$ and $D^+$ 
mesons are reconstructed. 
Figure~\ref{fig:ds_791}(right)
shows the 
reconstructed mass (difference from the nominal mass)
and decay time distributions. They find
$\tau(D^0) =  408.5 \pm  4.1    \, ^{+\, 3.5}_{-\,  3.4}$~fs,
$\tau(D^+) = 1033.6 \pm 22.1    \, ^{+\, \, 9.9}_{-\, 12.7}$~fs,
$\tau(D^+_s) =  486.3 \pm 15.0    \, ^{+\, 4.9}_{-\,  5.1}$~fs,
and the lifetime ratio of 
$\tau (D_s^+) / \tau(D^0) = 1.19 \pm 0.04$.

Each result establishes for the first 
time that the $D_s^+$ lifetime is
significantly longer than the $D^0$ lifetime.

\section{Bottom Hadron Lifetimes}\label{subsec:b_life}

\begin{figure}[t]
\mbox{ \epsfxsize=0.45\textwidth 
	\epsffile{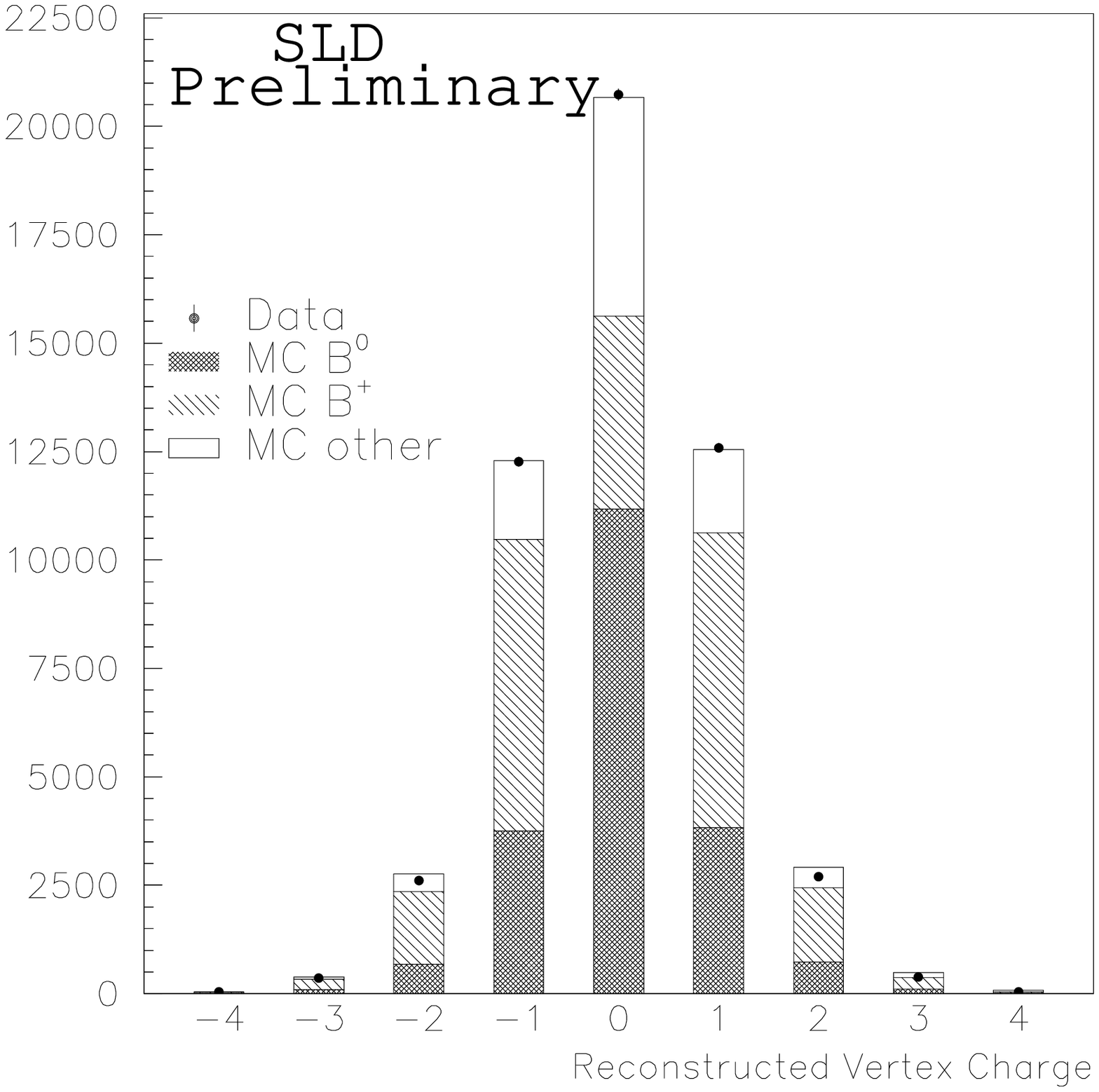}}
\vspace{-2mm}
\mbox{ \epsfxsize=0.52\textwidth 
	\epsffile{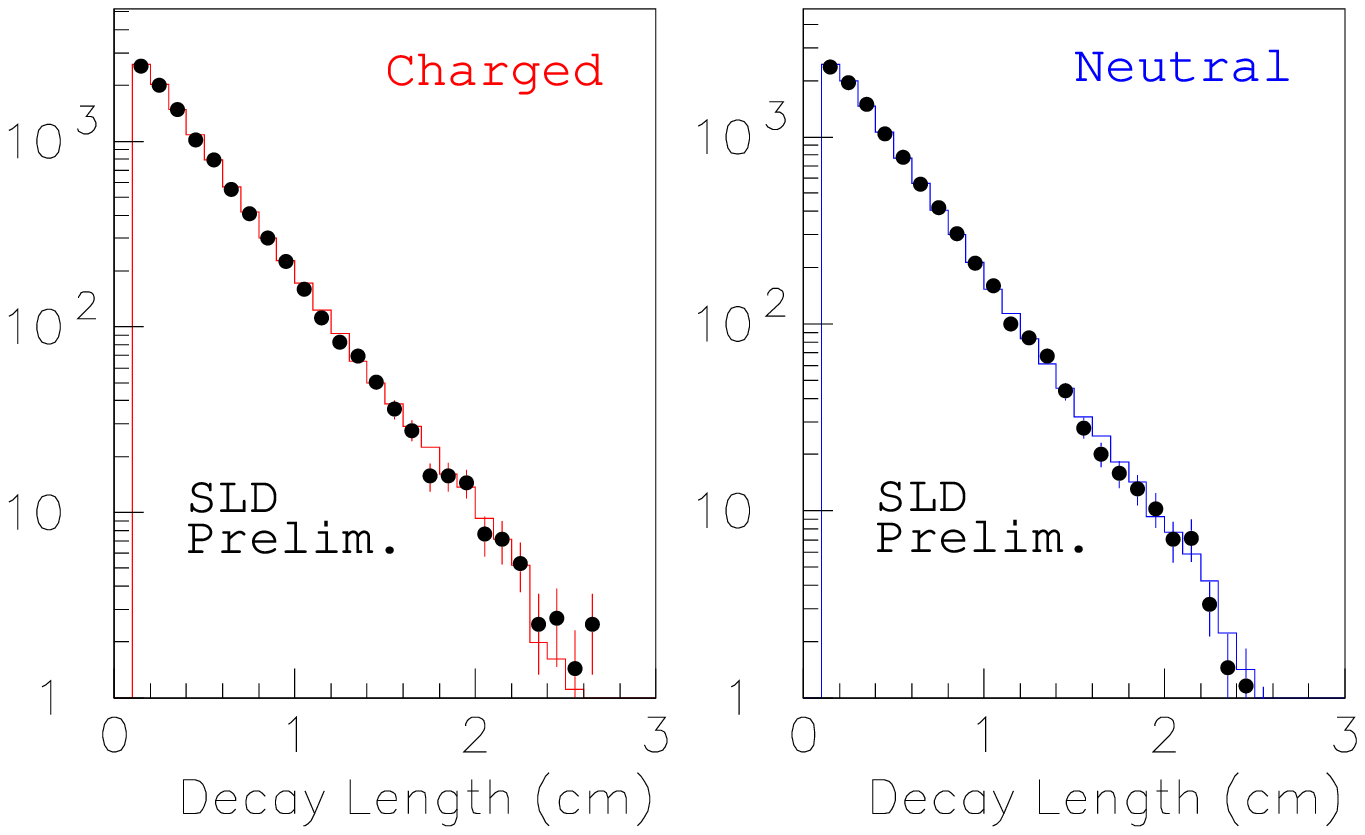}}

\caption{
$B^+$ and $B^0$ meson lifetime measurement by SLD.
Left: vertex charge distribution.
Right: 
decay length distributions for charged and neutral
$B$ hadron samples.
The plots show the 1997 -- 98 data 
corresponding to 350~k hadronic $Z^0$ events. 
}
\label{fig:sld}
\end{figure}

The bottom hadrons are expected to have smaller lifetime differences among the
different hadron species, of order 10\% at most~\cite{Bigi}.
This poses a challenge to experiments, because it requires great precision
to 
find and establish such small differences.
The $B$ hadron lifetime measurements performed thus far
can be classified into three broad
classes, using (a) inclusive $B$ reconstruction, 
(b) partial reconstruction (e.g.~semileptonic decays)
and (c) full reconstruction (e.g.~$J/\psi K^{(*)}$,  $D^{(*)} n\pi$).
Method (a) gives the largest statistics, but samples are less pure
in terms of isolating the various $B$ hadron species,
and the selection procedure introduces a bias in the
proper time distribution of the candidates 
that needs to be carefully accounted for.
Method (b) gives respectable statistics and sample purity.
Method (c) gives clean signature, perfect purity, and very precise
momentum estimate on an event-by-event basis, but suffers from
small branching fractions and thus low statistics. 
Here we describe two measurements, using method (a) and (b), respectively.


\begin{figure}  
\mbox{ \epsfxsize=0.48\textwidth 
	\epsffile{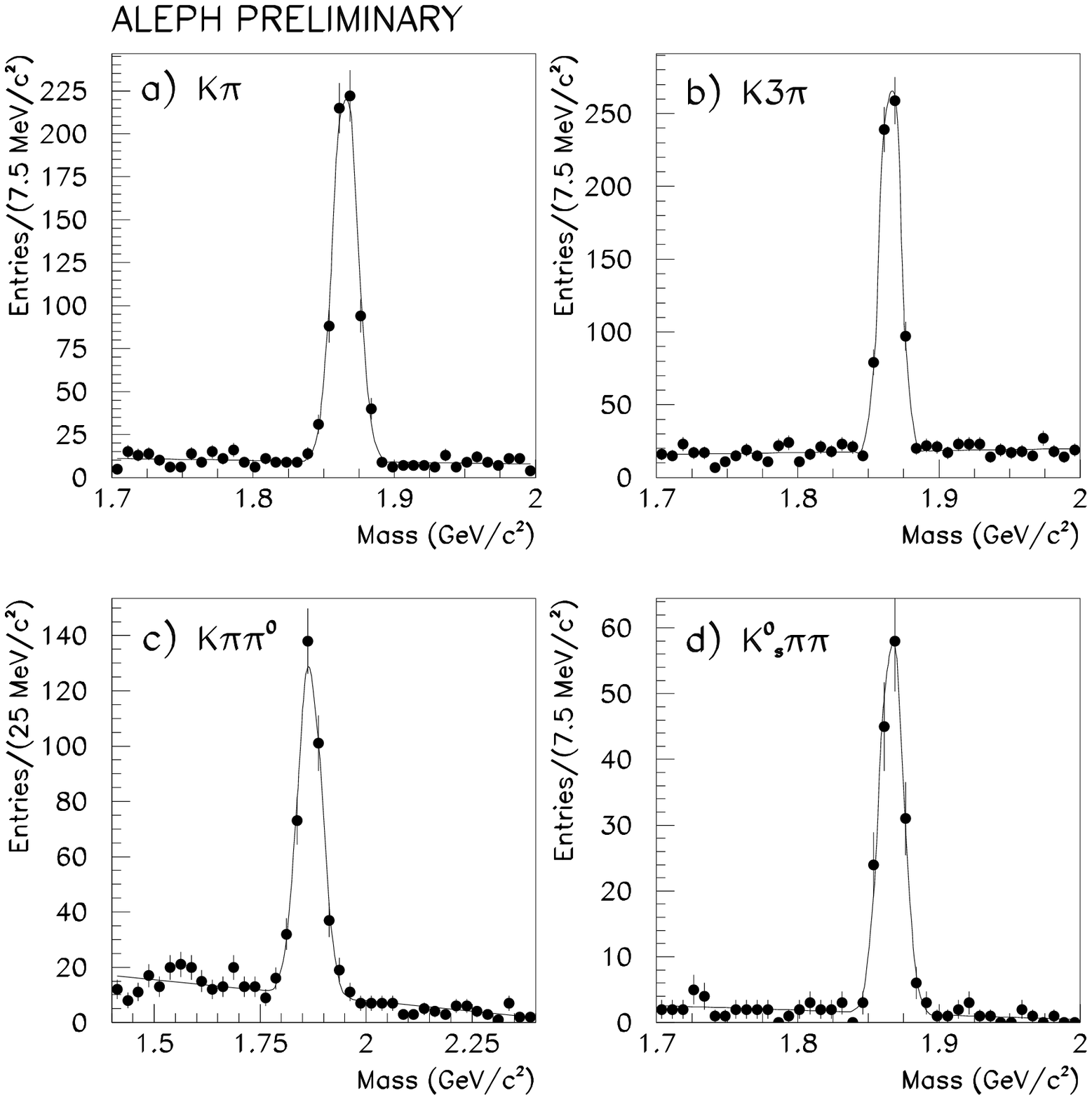}}
\vspace{-3mm}
\mbox{ \epsfxsize=0.47\textwidth 
	\epsffile{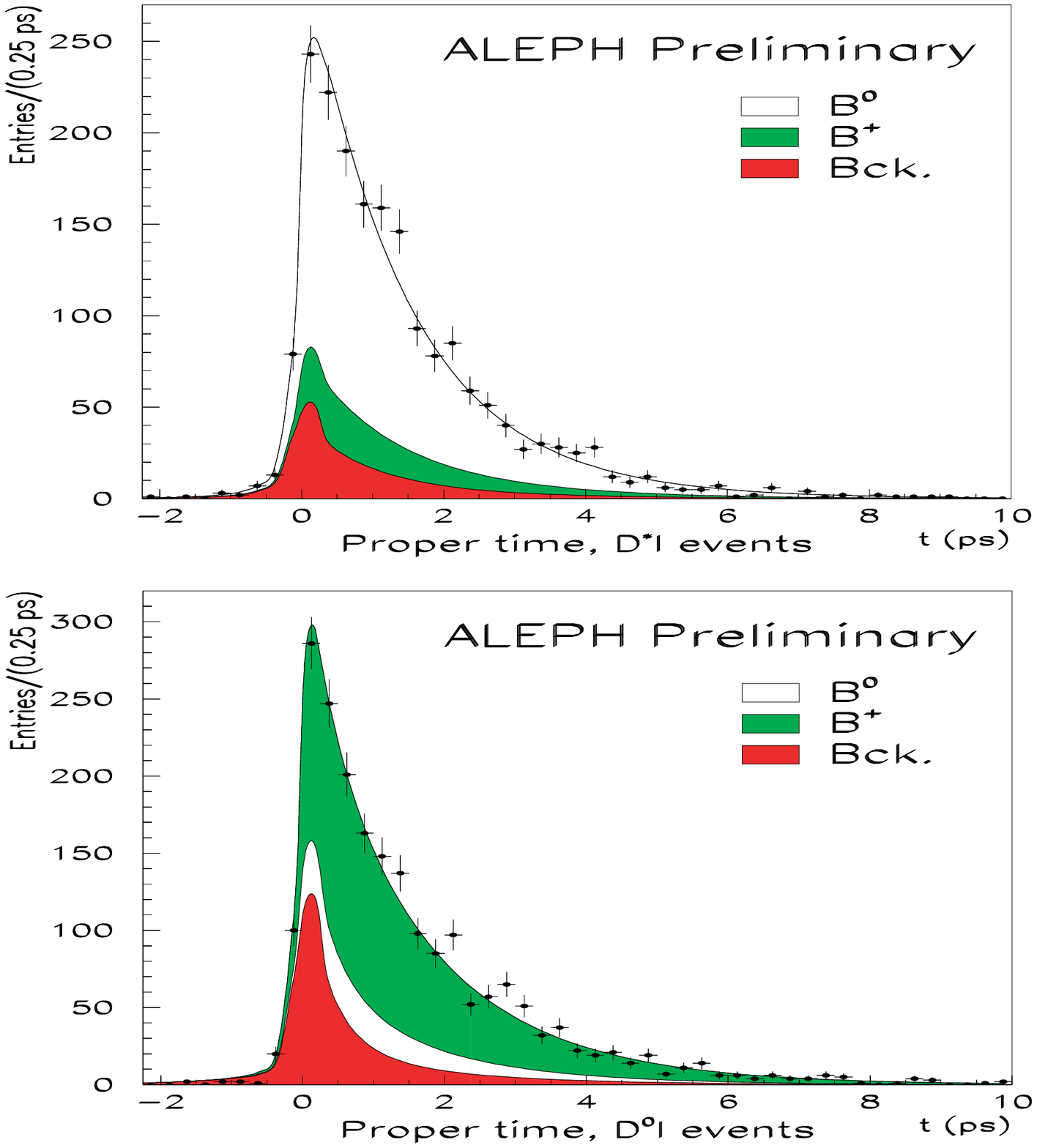}}
\caption{
$B^+$ and $B^0$ meson lifetime measurement by ALEPH.
Left: $D^0$ signal from 
$D^{*+} \rightarrow D^0 \pi^+$ associated 
with $\ell^-$.
Right: $B$ meson proper decay time distributions for 
$\ell^- D^{*+}$ and $\ell^- D^0$ samples.
}
\label{fig:aleph_life}
\end{figure}

SLD~\cite{SLD} uses inclusively reconstructed $B$ decays 
among 
550 k 
hadronic $Z^0$  decays. 
The net charge of tracks attached
to the secondary vertices is used to estimate and separate
the charge of 
parent $B$ hadrons. 
The purity is enhanced further with 
information from
the vertex mass, the beam polarization and the opposite hemisphere jet charge.
Figure~\ref{fig:sld} shows the distributions of the vertex charge
and the 
decay lengths for samples enriched in
charged and neutral $B$ hadrons.
They find 
$\tau(B^+)  =  1.623 \pm 0.020 \pm 0.034$~ps, 
$\tau(B^0)  =  1.565 \pm 0.021 \pm 0.043$~ps, 
and the lifetime ratio of
$\tau (B^+) / \tau(B^0) = 1.037  \, ^{+\, 0.025}_{-\, 0.024} \pm 0.024$.

ALEPH~\cite{aleph_life} has re-analyzed 4~M hadronic $Z^0$ decays
taken in 1991 -- 95 and measured 
the $B^+$ and $B^0$ meson lifetimes
using semileptonic decays 
${\overline B} \rightarrow \ell^- \bar \nu D^{*+} X$ 
(mostly ${\overline B}$$^0$) 
and ${\overline B} \rightarrow \ell^- \bar\nu D^0 X$ (mostly $B^-$).
The $D^{*+}$ meson is reconstructed in the $D^0 \pi^+$ mode,
and the $D^0$ meson is reconstructed in 
$K^- \pi^+$,    $K^- \pi^+ \pi^+ \pi^-$,
$K^- \pi^+ \pi^0$ or $K^0_S \pi^+ \pi^-$ mode.   
The charm signals are shown in Figure~\ref{fig:aleph_life}.
For the $\ell^- D^0$ pairs, 
the $D^0$ meson is reconstructed in 
$K^- \pi^+$,    
$K^- \pi^+ \pi^0$ or $K^0_S \pi^+ \pi^-$ mode, and 
decays coming from the $D^{*+}$ meson
are excluded. Similar signals  are observed. 
Distributions of the $B$ meson proper decay time
are also shown in Figure~\ref{fig:aleph_life}.
The measured lifetimes are
$\tau(B^+) =  1.646 \pm 0.056 \, ^{+\, 0.036}_{-\, 0.034} $~ps, 
$\tau(B^0) =  1.524 \pm 0.053 \, ^{+\, 0.035}_{-\, 0.032} $~ps, 
and the lifetime ratio is
$\tau (B^+) / \tau(B^0) = 1.080  \pm 0.062  \pm 0.024$.

These two results are shown as recent representative measurements.
There are many other measurements from various experiments,
as compiled by the LEP $B$ lifetime working group,
and more information can be found in Ref.~\cite{LEP_life}.
As of October 1999, the world average lifetimes of $B^+$ and $B^0$ 
mesons are 
$\tau(B^+) =  1.639 \pm 0.025$~ps, 
$\tau(B^0) =  1.553 \pm 0.029$~ps, 
and 
$\tau (B^+) / \tau(B^0) = 1.066 \pm 0.024$.
The 
value at the time of
the Hawaii Conference (March 1997) 
was
$\tau (B^+) / \tau(B^0) = 1.06 \pm 0.04$.
Namely, 
the precision is still improving, 
and we may be on the verge of observing a finite lifetime difference.

For the $B^0_s$ meson, 
a very small (of order 1\%) lifetime difference
from the $B^0$ meson
is expected.
The current world average value is 
$\tau (B^0_s) = 1.460 \pm 0.056$~ps, and 
$\tau (B^0_s) / \tau(B^0) = 0.94 \pm 0.07$. 
More precise measurements are necessary in order to see
whether
there is a lifetime difference of order 5\%,
or whether the lifetimes agree 
at 1\% level.

Another  interesting piece of physics 
is expected in the $B^0_s$ meson system.
The $B^0_s$ meson should mix with its
antiparticle to form the mass eigenstates $B^0_{s,L}$ and $B^0_{s,H}$,
where the subscripts $L$ and $H$ denote light and heavy states.
These two mass eigenstates could also have a substantial width difference 
$\Delta \Gamma$ of order 10\%.
A sizable width difference  is 
interesting for several reasons: 
(a) we should see it if it indeed exists,
(b) the ratio $\Delta m_s / \Delta \Gamma$ is independent of CKM 
matrix elements and is estimated to be large ($\sim -180$),
providing an indirect information on $\Delta m_s$,
(c) it may allow $CP$ studies using untagged decays~\cite{dunietz},
and (d) physics beyond the Standard Model 
should make $\Delta \Gamma$ smaller.

\begin{figure}  
\mbox{ \epsfxsize=0.48\textwidth 
	\epsffile{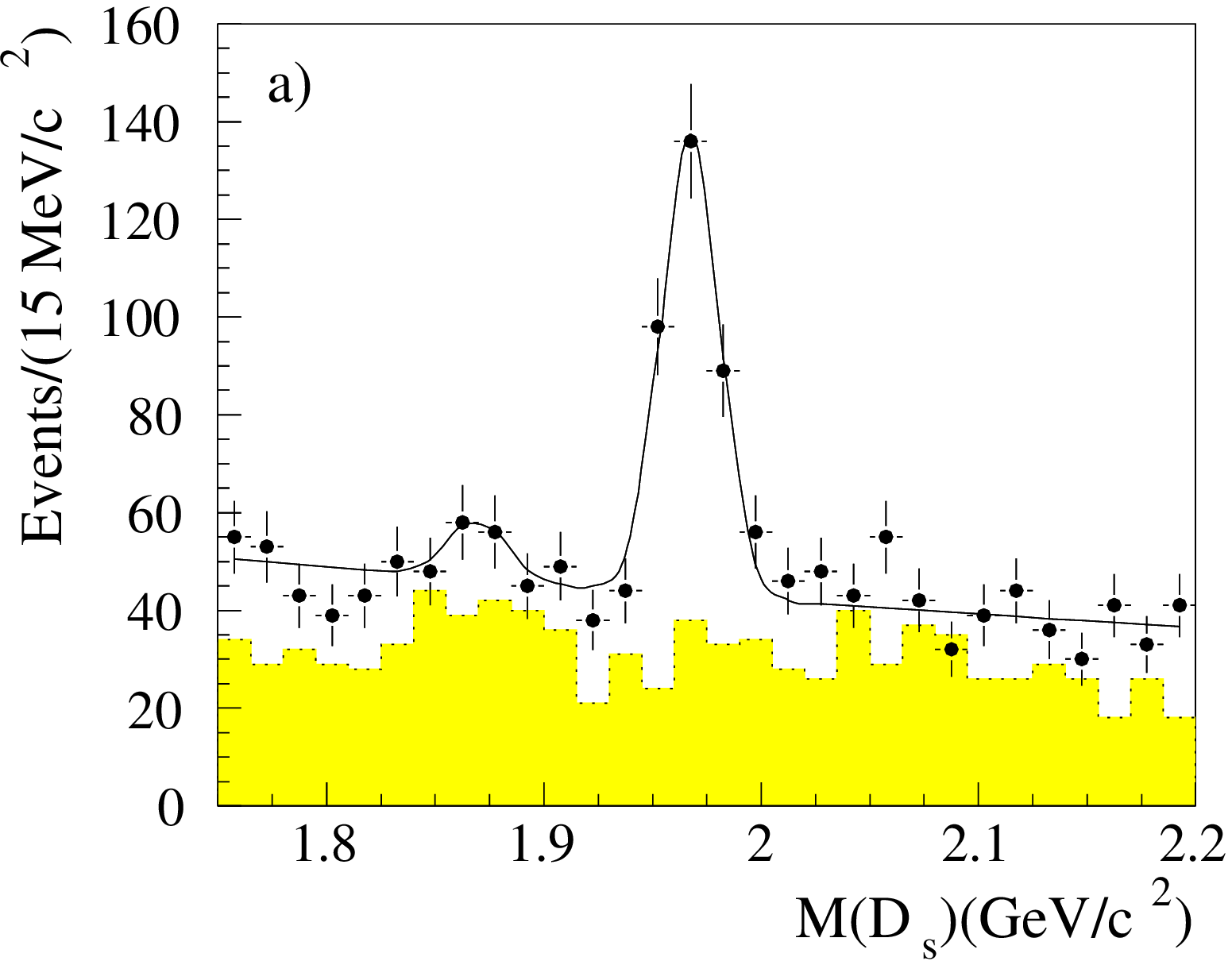}}
\vspace{-3mm}
\mbox{ \epsfxsize=0.47\textwidth 
	\epsffile{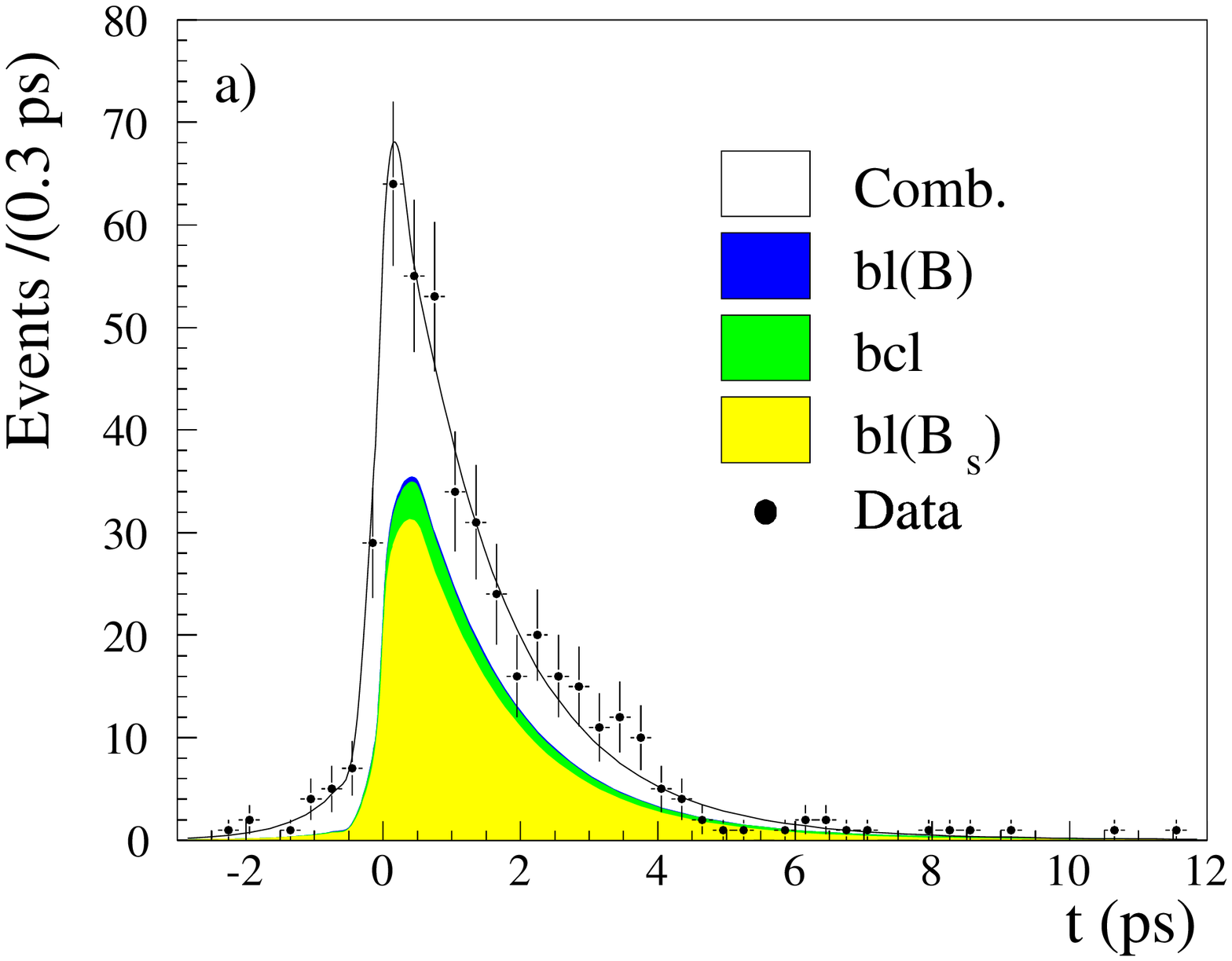}}
\caption{
$B^0_s$ meson lifetime measurement by DELPHI.
Left: $D^+_s$ signal associated with $\ell^-$.
Right: $B^0_s$ meson proper decay time distribution.
}
\label{fig:delphi_ds}
\end{figure}

The width difference $\Delta \Gamma$ manifests itself as
a difference in lifetimes when measured with 
flavor eigenstates (e.g. semileptonic decays)
and $CP$ eigenstates. 
Or, the decay time distributions of flavor eigenstates can be
fitted with two (long and short) components 
to look for a width difference.

DELPHI~\cite{delphi_dgamma} uses 3.5 M $Z^0$ decays and
the semileptonic decay
${\overline B}$$^0_s \rightarrow \ell^- \bar\nu D_s^+ X$.
The $D^+_s$ meson is reconstructed in 
$\phi \pi^+$, 
${\overline K}$$^{*0} K^{(*)+}$, 
$K^0_S K^+$, $\phi \pi^+ \pi^+ \pi^-$ or $\phi \pi^+ \pi^0$ mode,
as well as the semileptonic mode $\phi \ell^+ \nu$.
A total signal of 290 events is observed. 
Figure~\ref{fig:delphi_ds} shows the combined $D^+_s$ signal
(excluding the semileptonic mode)
and the $B^0_s$ decay time distribution.
They find 
$\tau(B^0_s) = 1.42  \, ^{+\, 0.14}_{-\, 0.13} \pm 0.03$~ps, 
and 
$\Delta \Gamma / \Gamma < 0.46$. 
DELPHI also uses 
${\overline B}$$^0_s \rightarrow D_s^+ h^- X$ mode 
and find 
$\tau(B^0_s) = 1.49  \, ^{+\, 0.16}_{-\, 0.15} \, ^{+\, 0.07}_{-\, 0.08}$~ps
and 
$\Delta \Gamma / \Gamma < 0.58$. 
Other $\Delta \Gamma$ searches include
an L3 analysis~\cite{L3_dgamma} using inclusive $B$ reconstruction, 
yielding 
$\Delta \Gamma / \Gamma < 0.67$ 
and a CDF analysis~\cite{CDF_dgamma} using
$\sim 600$ $\ell^- D^+_s$ decays,
yielding
$\tau(B^0_s) = 1.36  \pm 0.09  \, ^{+\, 0.06}_{-\, 0.05}$~ps
and $\Delta \Gamma / \Gamma < 0.83$. 
These limits
on the width difference,
all at 95\% CL,
have been
obtained by examining decay time distributions and
looking for two components.

The $B^0_s$ lifetime is also measured 
using modes that are nearly $CP$ eigenstates.
ALEPH~\cite{aleph_ds_ds} uses the 
$B^0_s/{\overline B}$$^0_s \rightarrow D_s^{(*)+} D_s^{(*)-}
\rightarrow \phi \phi X$
signature with $32 \pm 17$ signal events,
and finds
$\tau(B^0_s) = 1.42   \pm 0.23 \pm 0.16$~ps.
CDF~\cite{cdf_psi_phi}
uses 
$B^0_s/{\overline B}$$^0_s \rightarrow J/\psi \, \phi$  
decays 
with $58 \pm 12$ signal events, 
and finds
$\tau(B^0_s) = 1.34   \, ^{+\,0.23} _{-\, 0.19} \pm 0.05$~ps.
These decays are not in general pure $CP$ eigenstates.
CDF  
performs a transversity analysis~\cite{cdf_pol} 
of the $J/\psi \, \phi$ 
decays and finds the fraction of the $P$-wave decays ($= CP$ odd)
to be $\Gamma_{\perp} / \Gamma = 0.229 \pm 0.188 \pm 0.038$.
For a similar decay mode $B^0 \rightarrow J/\psi K^{*0}$, 
CDF~\cite{cdf_pol}  
and CLEO~\cite{cleo_pol} 
find the fraction
to be 
$0.126  \, ^{+\,0.121} _{-\, 0.098} \pm 0.028$
and $0.16 \pm 0.08 \pm 0.04$,
respectively.

The above lifetime values may be compared with 
the world average 
$B^0_s$ lifetime measured with
flavor eigenstates, 
$\tau(B^0_s) = 1.467 \pm 0.058$~ps.
This is smaller, albeit large uncertainties,
as expected for $CP$-even states.
The ALEPH analysis quotes
$\Delta \Gamma / \Gamma = 0.24 \pm 0.35$
assuming pure $CP$-even composition with
$\tau(B^0_s) = 1.61 \pm 0.10$~ps (PDG 1996).


Since I have reached the page limit, 
let me conclude   
here by saying that new data expected 
in the near future
from CLEO-III, Belle, BaBar and Tevatron Run-II 
should vastly improve the kinds of measurements described here.

\section*{Acknowledgments}
I wish to thank Professor George Hou,
my old friend Augustine Chen, 
and the other members 
of the local organizing committee for a very nice 
conference. 
Also, I would like to thank the experiments
for their physics results
contributing to this talk.
In particular, the following people 
have provided me with great help in preparing the talk
and this manuscript:
D.~Jaffe, 
J.~Slaughter, 	
K.~Baird, D.~Jackson, 
D.~Abbaneo, 
P.~Gagnon, R.~Hawkings,  
S.~Gentile, S. Blyth  
and A.~B.~Wicklund.
Professor Joseph Kroll (Penn) has provided brilliant ideas
and incredible inspiration as always.


\end{document}